\documentclass[twocolumn,prl,showpacs,nofootinbib,superscriptaddress]{revtex4}
\usepackage{latexsym}
\usepackage{dcolumn}
\usepackage{epsfig}
\RequirePackage{graphicx}

\newcommand{\ba}{\begin{eqnarray}}

\newcommand{\ea}{\end{eqnarray}}
\newcommand{\be}{\begin{equation}}
\newcommand{\ee}{\end{equation}}

\newcommand{\eq}[1]{Eq.\,(\ref{#1})}

\newcommand{\lessabout}{\raisebox{-.6ex}{\ $\stackrel{<}{\sim }$\ }}

\newcommand{\alphabold}{\mbox{\small\boldmath $\alpha$}}
\newcommand{\xbold}{\mbox{\boldmath $x$}}

\newcommand{\delchisq}{\Delta \chi^2_i(x_i;\alphabold)}
\newcommand{\delchi}{\Delta \chi^2_i}

\newcommand{\delchimax}{{\delchi}_{\rm max}}

\def\bea{\begin{eqnarray}} 
\def\eea{\end{eqnarray}} 
\begin{document}

\preprint{ANL-HEP-PR-07-13}

\title{Analytic Expression for the Joint $x$ and $Q^2$ Dependences of \\
the Structure Functions of Deep Inelastic Scattering}
\author{Edmond~L.~Berger}
\affiliation{High Energy Physics Division,
Argonne National Laboratory,
Argonne, Illinois 60439}
\author{M.~M.~Block}
\affiliation{Department of Physics and Astronomy, Northwestern University, 
Evanston, IL 60208}
\author{Chung-I Tan}
\affiliation{Physics Department, Brown University, Providence, RI 02912} 
\date{\today}

\begin{abstract}
We obtain a good analytic fit to the joint Bjorken-$x$ and $Q^2$ dependences 
of ZEUS data on the deep inelastic structure function $F_2(x, Q^2)$. At fixed 
virtuality $Q^2$, as we showed previously, our expression is an expansion 
in powers of $\ln(1/x)$ that satisfies the Froissart bound. Here we 
show that for each $x$, the $Q^2$ dependence of the data is well described 
by an expansion in powers of $\ln Q^2$.  The resulting analytic expression 
allows us to predict the logarithmic derivatives 
${({\partial}^n F_2^p/{{(\partial\ln Q^2}})^n)}_x$ for $n = 1,2$ and to 
compare the results successfully with other data.  We extrapolate the proton 
structure function 
$F_2^p(x,Q^2)$ to the very large $Q^2$ and the very small $x$ 
regions that are inaccessible to present day experiments and contrast our 
expectations with those of conventional global fits of parton distribution 
functions. 
\end{abstract}


\maketitle

{\em Introduction.}  The ability to predict cross sections at very high energies, 
whether at the CERN Large Hadron Collider or in ultra high energy cosmic ray 
interactions, depends critically on the reliability of extrapolations from 
current measurements into regions of much greater virtuality ($Q^2$) of the elementary 
scattering processes, and to much smaller values of the fractional longitudinal 
momentum $x$ carried by the parton constituents of the hadrons.  Most high energy 
predictions are expressed in terms of convolutions of elementary hard-scattering cross 
sections with parton distribution functions (PDFs).  The quantitative reliability of 
these predictions relies 
on the $x$ and $Q^2$ dependences embodied in the universal parton distribution 
functions extracted from global analyses in perturbative quantum chromodynamics 
(QCD) of data at lower energies. Important in these global analyses 
are data from deep-inelastic lepton scattering (DIS) and other reactions, but 
equally critical are the analytic functional forms assumed for the $x$ dependence 
of parton distribution functions at small $x$. 

In an earlier paper~\cite{bbt}, we analyze the $x$ dependence of the DIS proton 
structure functions $F^p_2(x, Q^2)$.  We begin with the assumption that the 
$x$ dependence at extremely small x should manifest a behavior consistent 
with saturation of the Froissart bound on hadronic total cross 
sections~\cite{froissart}, as is satisfied by data on 
$\gamma p$, $\pi^{\pm} p$, and $\bar pp$ and $pp$ interactions~\cite{IgiBlock}.  
This bound~\cite{froissart}, derived from analyticity and unitarity, demands 
that $F_2^p(x,Q^2)$ grow no more rapidly than $\ln^2(1/x)$ at very small $x$ 
for all values of $Q^2$. Over the ranges of $x$ and $Q^2$ for which DIS data 
are available, we then show that a very good fit to the $x$ dependence of ZEUS 
data 
~\cite{ZEUS} is obtained for 
$x \le x_P = 0.09$ and ${Q^2\over x}\gg m^2$ with the expression  
\ba
F_2^p(x,Q^2)=(1-x)&\!\!\!\times&\!\!\!\!\Big\{{F_P\over{1-x_P}}+A(Q^2)\ln\left[\frac {x_P}{x}\frac{1-x}{1-x_P}\right] \nonumber\\
& \!\!\!+ & 
B(Q^2)\ln^2\left[\frac {x_P}{x}\frac{1-x}{1-x_P}\right]\Big\}. \label{Fofx} 
\ea
Our fits to DIS data~\cite{ZEUS} at 24 values of 
$Q^2$ cover the wide range $0.11\le Q^2\le 1200$ GeV$^2$.  The value $x_P = 0.09$ is 
a scaling point~\cite{bbt} such that the curves for all $Q^2$ pass through the 
point $x = x_P$, at which $F_2(x_P, Q^2) = F_P \sim 0.41$, constraining all of 
the fits.

In this note, we extend our analysis by making a {\em joint fit of both} the 
$x$ and $Q^2$ dependences of the ZEUS~\cite{ZEUS} data on $F_2^p(x, Q^2)$.  The 
analytic expression we derive for the $x$ and $Q^2$ dependences allows us to 
compute the logarithmic partial derivatives 
${({\partial}^n F_2^p(x,Q^2)/{(\partial\ln Q^2})^n)}_x$ for $n = 1, 2$. We obtain 
excellent agreement when comparing our predictions for the first derivative ($n = 1$) 
with H1 data~\cite{H1}.  We offer predictions for the 
second derivative ($n = 2$). Only 8 parameters---two 
of which are the scaling value $F_P =0.41$ at the scaling point $x_P=0.9$---are needed 
to fit the joint $x$ and $Q^2$ dependences.  Our expression allows us to extrapolate 
to very low values of $x$, well beyond the experimental range presently accessible. We  
obtain cross sections for ultra high energy cosmic ray neutrino reactions 
that are a factor of $\sim 5$ smaller than those based on extrapolations of conventional 
parton distribution functions.     

{\em Joint Fit.}  In \eq{Fofx}, $F_2^p(x, Q^2)$ is written as a sum of terms 
that are factorizable as functions of $Q^2$  times functions of $x$ that 
satisfy the Froissart bound.  In this paper we  discuss our fit for the 
functions $A(Q^2)$ and $B(Q^2)$.

To parameterize the dependence on $Q^2$ at fixed $x$, we assume an expansion in 
powers of $\ln Q^2$, generally 
consistent with and motivated by the $\ln Q^2$ variation expected in QCD.  We note, 
moreover, that the H1 collaboration~\cite{H1} determined that, for  fixed $x$, the $Q^2$ 
dependence of $F_2^p(x,Q^2)$ is reproduced by the form 
$F_2^p(x,Q^2)=\alpha_0(x)+\alpha_1(x)\ln(Q^2)+\alpha_2(x)\ln^2(Q^2)$.  We therefore 
expand the functions $A(Q^2)$ and $B(Q^2)$ as 
\ba 
    A(Q^2)&=&a_0+a_1\ln Q^2 +a_2\ln^2 Q^2, \nonumber \\ 
    B(Q^2)&=&b_0+b_1\ln Q^2 +b_2\ln^2 Q^2,  \label{AB}
\ea
terminating these phenomenological expansions at the quadratic level.  

We fit simultaneously the $x$ dependence and the $Q^2$ dependence of the 
data ($Q^2$ is expressed in GeV$^2$ throughout).  We determine the 6 real constants 
$a_0,a_1,a_2,b_0,b_1$ and $b_2$ in \eq{AB} using the Sieve 
algorithm~\cite{sieve}, by minimizing the squared Lorentzian,
\be
\Lambda^2_0(\alphabold;\xbold)\equiv\sum_{i=1}^N\ln\left\{1+0.179\delchisq\right\},\label{lambda0}
\ee
where $\chi^2(\alphabold;\xbold)\equiv\sum_{i=1}^N\delchisq $, $\delchisq\equiv 
\left(\left[\bar y_i(x_i;\alphabold)-y_i(x_i)\right]/\sigma_i\right)^2$, $\alphabold$ 
is the parameter space vector, and $\bar y_i(x_i;\alphabold)$ is the theoretical value 
of the measured $y_i$ at $x_i$, with measurement error $\sigma_i$. Using a $\delchimax$ 
cut of 6, we find (see Table \ref{fitted}) a final corrected 
$\chi^2/$d.f.=1.09, for 169 degrees of freedom (d.f.), a reasonable fit.  The data used 
are 24 ZEUS data sets, with 
$Q^2=$0.11, 0.25, 0.65, 2.7, 3.5, 4.5, 6.5, 8.5, 10, 12, 15, 18, 22, 27, 35, 45, 70, 90, 120, 200, 250, 450, 800 and 1200 GeV$^2$. 
The Sieve algorithm eliminated 8 outlier points (from a total of 183) which had a 
$\chi^2$ contribution of 63.5.

\begin{figure}[h,t] 
\begin{center}
\mbox{\epsfig{file=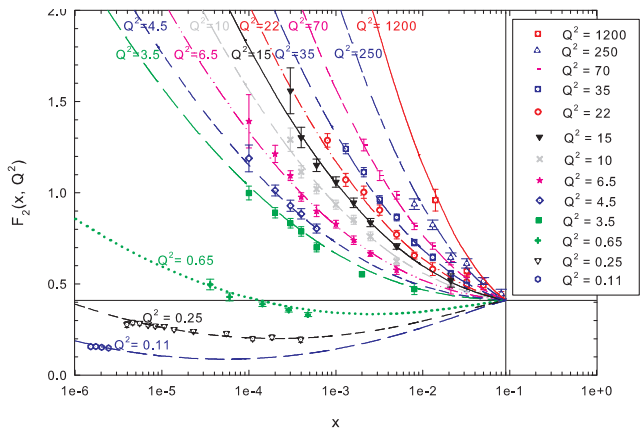,width=3.4in%
,bbllx=0pt,bblly=390pt,bburx=450pt,bbury=700pt,clip=%
}}
\end{center}
\caption[]{ 
Fits to the proton structure function data, $F_2^p(x, Q^2)$ vs. $x$, for 13 values of $Q^2$.  
The data are from the ZEUS collaboration~\cite{ZEUS}. The curves show 13 of our 28 global fits 
whose parameters are given in Table \ref{fitted}.
The vertical and horizontal straight lines intersect at the 
scaling point $x_{\rm P}=0.09,F_2^p(x_{\rm P})=0.41$.  \label{fig:Fvsx}}
\end{figure}

The quality of our fit to the $x$ and $Q^2$ dependences of the data for 
$x\le x_P,\quad {Q^2\over x}\gg m^2$ is shown in Fig.~\ref{fig:Fvsx}, a 
representative plot of 13 of the data sets.  
It shows that the fit is excellent over 
the large $Q^2$ and $x$ ranges of the ZEUS~\cite{ZEUS} data, fitting data equally well at 
$x\sim 10^{-6}$ for $Q^2=0.11$~GeV$^2$ as at $x\sim 10^{-2}$ for $Q^2=1200$ GeV$^2$, even 
with constraint that {\em all} curves must pass through the common scaling point 
$x_P=0.09,\ F_P=0.41$.  The 6 fit parameters are given in Table \ref{fitted}, along with 
their statistical errors. 
\begin{table}[h]                   
%
\begin{center}
\def\arraystretch{1.2}            
     \caption{\label{fitted}\protect\small Results of a 6-parameter fit to $F_2^p(x,Q^2)$ structure function data\cite{ZEUS} using the $x$ 
and $Q^2$ behaviors of \eq{Fofx} and \eq{AB}, with $Q^2$ in GeV$^2$.   The renormalized 
$\chi^2_{\rm min}$ per degree of freedom, taking into account the effects of the 
$\delchimax=6$ cut~\cite{sieve}, is given in the row labeled 
${\cal R}\times\chi^2_{\rm min}$/d.f. The errors in the fitted parameters are 
multiplied by the appropriate $r_{\chi2}$\cite{sieve}. }

\begin{tabular}[b]{|l||c||}     
	\multicolumn{1}{l}{Parameters}&\multicolumn{1}{c} {Values}\\
\hline
      $a_0$&$-5.381\times 10^{-2}\pm 2.17\times 10^{-3}$ \\ 
      $a_1$&$2.034\times 10^{-2}\pm 1.19\times 10^{-3}$\\ 
      $a_2$&$4.999\times 10^{-4}\pm 2.23\times 10^{-4}$\\
\hline
	$b_0$ &$9.955\times 10^{-3}\pm 3.09\times 10^{-4}$\\
      $b_1$&$3.810\times 10^{-3}\pm 1.73\times 10^{-4}$\\
      $b_2$&$9.923\times 10^{-4}\pm 2.85\times 10^{-5}$ \\ 
	\cline{1-2}
     	\hline
	\hline
	$\chi^2_{\rm min}$&165.99\\
	${\cal R}\times\chi^2_{\rm min}$&184.2\\ 
	d.f.&169\\
\hline
	${\cal R}\times\chi^2_{\rm min}$/d.f.&1.09\\
\hline
\end{tabular}
\end{center}
\end{table}
\def\arraystretch{1}  

\begin{figure}
\begin{center}
\mbox{\epsfig{file=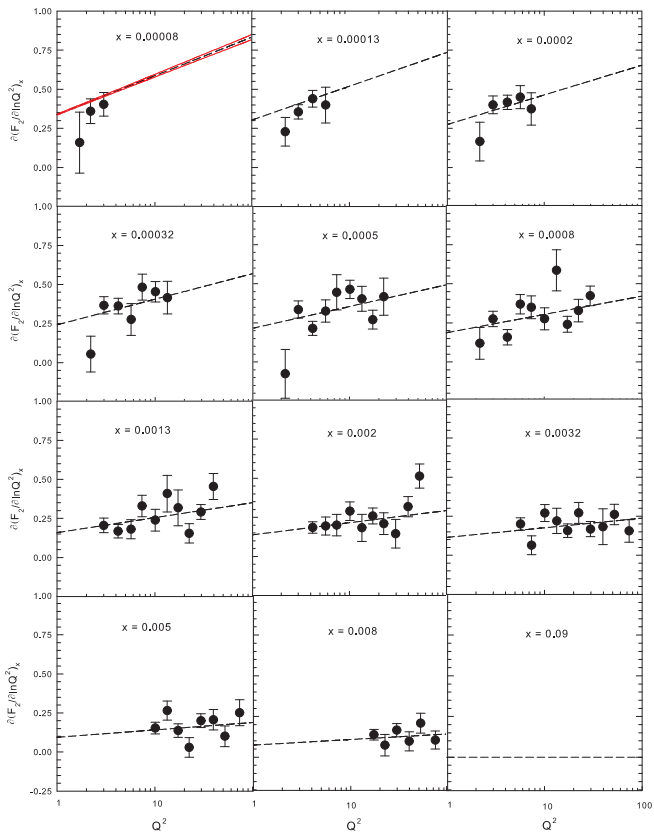,width=3.4in%
,bbllx=0pt,bblly=0pt,bburx=540pt,bbury=695pt,clip=%
}}
\end{center}
\caption[]{
A plot of the derivative $\partial {F_2^p}_x(x,Q^2)/{\partial \ln (Q^2)}$ vs. $Q^2$, in GeV$^2$, 
for selected values of $x$, compared to data from the H1 collaboration~\cite{H1}.  The exterior 
lines for $x=0.00008$ are the error bands associated with the parameter uncertainties of the 
coefficients of Table \ref{fitted}.}
\label{fig:dF}
\end{figure}

{\em Evaluation of $\partial {F_2^p}_x(x,Q^2)/\partial \ln (Q^2)$.}  Differentiating \eq{Fofx} 
with respect to $\ln(Q^2)$, we obtain 
\ba
{\partial {F_2^p}_x(x,Q^2)\over{\!\!\!\!\!\!\!\!\!\!\partial \ln (Q^2)}}&=&(1-x)\times\nonumber\\
&&\!\!\!\!\!\!\!\!\!\!\!\!\!\!\!\!\!\!\!\!\!\!\!\!\!\!\left\{\left(a_1+2a_2\ln Q^2\right)\ln\left[\frac {x_P}{x}\frac{1-x}{1-x_P}\right]+\right.\nonumber\\
&&\!\!\!\!\!\!\!\!\!\!\!\!\!\!\!\!\!\!\!\!\!\!\left.\left(b_1+2b_2\ln Q^2\right)\ln^2\left[\frac {x_P}{x}\frac{1-x}{1-x_P}\right]\right\},\label{dF}
\ea
an expression valid  for $x\le x_P$ and $Q^2/x \gg m^2$. We show a plot of 
$\partial {F_2^p}_x(x,Q^2)/{\partial \ln (Q^2)}$ in Fig. \ref{fig:dF} 
for a set of values of $x$, compared to the values 
measured by the H1 collaboration~\cite{H1}.  
We emphasize that the theoretical values have been constrained by ZEUS~\cite{ZEUS} data 
alone, and that they are a {\em prediction} of the H1 results, {\em not} a fit to these data.  
We see from Fig. \ref{fig:dF} that our predictions based on the ZEUS data 
are in fine agreement with the normalization and slope of the H1 results.

{\em Curvature.}  The curvature,  defined as one-half of the second derivative, is given by
\ba
{\rm curvature}=\frac{1}{2}{{\partial}^2 {F_2^p}_x(x,Q^2)\over{\!\!\!\!\!\!\!\!\!\!\partial {\ln (Q^2)^2}}}
&=&(1-x)\times\nonumber\\
&&\!\!\!\!\!\!\!\!\!\!\!\!\!\!\!\!\!\!\!\!\!\!\!\!\!\!\!\!\!\!\!\!\!\!\!\!\!\!\!\!\!\!\!\!\!\!\!\!\!\!\!\!\!\!\!\!\!\!\!\!\!\!\!\!\!\!\!\!\!\!\!\!\!\!\!\!\left\{\!a_2\ln\!\left[\frac {x_P}{x}\frac{1-x}{1-x_P}\right]\!+\!\right.\left.b_2\ln^2\left[\frac {x_P}{x}\frac{1-x}{1-x_P}\right]\!\right\}\!.\label{dF2}
\ea
The form of \eq{dF2} indicates that our 
curvature is independent of $Q^2$, a consequence of the fact that we truncate the expansions in 
\eq{AB} at the quadratic level.  With our parameterization, the curvature grows like $\ln^2(1/x)$ 
as $x$ decreases.  The signs and magnitudes of $a_2$ and $b_2$ determine the sign (positive/negative) of the 
curvature.  In our case, the curvature is positive, and it increases as $x$ decreases, features that are also 
true in next-to-leading order perturbative QCD~\cite{Gluck:2006pm}.  We note, however, that we do not impose or 
employ QCD evolution in our work.  Our predictions are based entirely on our fit to data on 
$F_2^p(x, Q^2)$.  The results of our calculation of curvature are in good agreement with the data 
shown in Ref.~\cite{Haidt:2004ck}.  We remark that curvature is defined somewhat differently in 
Ref.~\cite{Haidt:2004ck} as the second derivative with respect to $\rm{log}_{10}(1 + Q^2/Q_o^2)$, 
instead of with respect to $\ln(Q^2)$.  

\begin{figure}[b,t] 
\begin{center}
\mbox{\epsfig{file=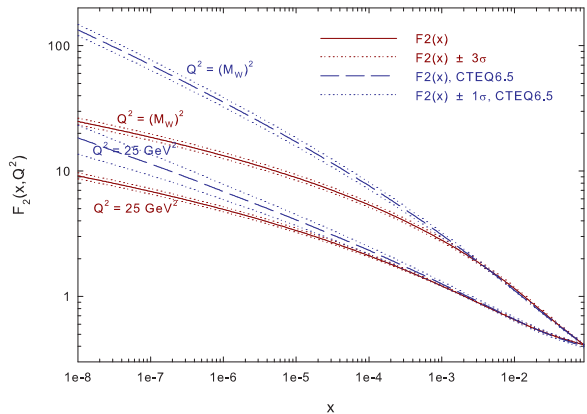,width=3.3in%
,bbllx=0pt,bblly=0pt,bburx=450pt,bbury=330pt,clip=%
}}
\end{center}
\caption[]{
A plot of $F_2^p(x, Q^2)$ vs. $x$ at $Q^2 = 25$~GeV$^2$ and $Q^2 = M_W^2$, where $M_W$ is the mass of the $W$ 
boson, along with results based on the CTEQ6.5M parton distribution functions~\cite{Tung:2006tb}.}
\label{fig:verysmallx}
\end{figure}

{\em Extrapolation to Very Small $x$.}  In Fig.~\ref{fig:verysmallx}, we present our 
calculation of $F_2^p(x, Q^2)$ as a function of 
$x$ for the choices $Q^2 = 25$~GeV$^2$ and $Q^2 = M_W^2$, where $M_W$ is the mass of the 
intermediate $W$ boson.  The scale $Q = M_W$ is of interest at hadron colliders where it 
characterizes electroweak processes.  It is also the relevant scale in charged-current high 
energy neutrino interactions~\cite{Gandhi:1998ri} where the $W$ boson propagator limits 
momentum transfers to $Q^2 \sim M_W^2$.  
 
We contrast our expectations with evaluations of $F_2^p(x, Q^2)$ based on the CTEQ6.5 set 
of parton distribution functions~\cite{Tung:2006tb}. In our case, the uncertainty bands 
represent a $\pm 3$ 
standard deviation variation of our parameters, whereas in the CTEQ6.5 case the bands are 
obtained from the 40 eigenvector sets that encapsulate the uncertainties of their PDFs.  

Inclusive $W$ boson production at hadron colliders serves as an independent probe of 
the $x$ dependence of quark and antiquark densities at $Q = M_W$, sensitive
to values of $x \sim (M_W/\sqrt s) e^{-y}$, where $s$ is the square of the center-of-mass energy  
and $y$ is the rapidity of the $W$ boson~\cite{Berger:1988tu}.  At the Fermilab Tevatron, 
with $\sqrt s = 1.96$~TeV, values of $x$ down to $2 \times 10^{-3}$ are probed for $y \sim 3$, well 
within the range in which our expectation and that of CTEQ6.5 show agreement in  Fig.~\ref{fig:verysmallx}. 

The magnitude and $x$ dependence of the CTEQ6.5 and our 
calculations agree quite well over the range $10^{-3}< x < 0.1$.  Both approaches also show 
the same dependence on $Q^2$ in this region of $x$.  The agreement is expected since both fit 
data that are limited to this range of $x$ at large $Q^2$.  The agreement also shows that the 
logarithmic expansion we use to describe $x$ dependence and the inverse power behavior of the 
CTEQ form cannot be distinguished numerically over the finite range $10^{-3}< x < 0.1$.  However, 
the two expectations clearly diverge considerably when extrapolated to values of $x$ as low as 
$10^{-8}$.  In the absence of new physics effects in the small-$x$ region, the 
saturation of the Froissart bound embodied in our fit to data at available energies yields 
a robust extrapolation into the region of very small $x$.

In the parton model, the decomposition of the structure function
$F_2^p(x, Q^2)$ at very small
$x$ is dominated by the sea quark $q(x, Q^2)$ and sea antiquark 
$\bar{q}(x, Q^2)$ densities.
Although we do not decompose our $F_2^p$ into parton distributions, this 
dominance allows us to
conclude from Fig.~\ref{fig:verysmallx} that our sea quark (and antiquark) 
distributions
will be $\sim 5$ times smaller than those in CTEQ6.5 at $x \sim 10^{-8}$ 
and $Q^2\sim M_W^2$.  In
ultra high energy charged-current neutrino 
interactions~\cite{Gandhi:1998ri},
$x \sim M_W^2/2mE_{\nu}$, and the cross section on (isoscalar) nucleons is 
proportional  to the sum 
of the up and down antiquark distributions, $\bar{u}(x,M_W^2) + \bar{d}(x,M_W^2)$.  
From 
Fig.~\ref{fig:verysmallx}, we see that the expected cross section at $x 
\sim 10^{-8}$ will
be  $\sim 5$ times smaller than that based on CTEQ6.5~\cite{comment}, serving to define a range of uncertainty for estimated
rates, with our expectation coming in on the low side.

{\em Summary.} The Bjorken-$x$ dependence of the DIS proton structure function $F_2^p(x, Q^2)$ measured 
by the ZEUS collaboration is consistent with a $\ln^2 (1/x)$ dependence at small values of $x$, compatible 
with saturation of the Froissart bound at each value of $Q^2$.  We parameterize successfully the joint $x$ 
and $Q^2$ dependences of $F_2$ for $x\le x_P \sim 0.09$ and ${Q^2\over x}\gg m^2$, using the compact 
factorized expression in \eq{Fofx}, 
with the $Q^2$ variation expressed in \eq{AB}.  Our analytic expression has only 6 parameters 
(plus the scaling point $x_P$ and the value $F_2^p(x_P)$ at the scaling point).  We compute the first and 
second derivatives of 
$F_2^p(x, Q^2)$ with respect to $\ln Q^2$ at small $x$.  Our predictions of these quantities are 
in good agreement with the measurements by 
the H1 collaboration~\cite{H1}.  We extrapolate our expression for $F_2^p(x, Q^2)$ down to the very 
small value $x = 10^{-8}$ and compare our expectations to those based on the CTEQ6.5M parton 
distributions~\cite{Tung:2006tb}.   

Under the assumption that the Froissart bound applies to the virtual photon cross section 
$\sigma(\gamma^*p)$, a $\ln^2(1/x)$ behavior is as singular as is allowed for the very small $x$ 
behavior of $F_2^p(x, Q^2)$.  However, it is difficult to reconcile a $\ln^2(1/x)$ behavior at 
very small $x$ for $F_2^p(x, Q^2)$ [and for the gluon distribution $g(x, Q^2)$] at all $Q^2$ with 
the Dokshitzer Gribov Lipatov Altarelli Parisi (DGLAP) evolution equation~\cite{dglap} at 
next-to-leading order in QCD, owing to the singular nature of parton splitting functions at small 
$x$.  Global analyses of parton distribution functions based on DGLAP evolution, such as CTEQ6.5, 
begin with the assumption of an inverse power behavior for the small $x$ dependences of the 
quark, antiquark, and gluon densities, behavior that is more singular than is allowed by the Froissart 
bound.  The assumed inverse-power behavior leads to the very different expectations shown in 
Fig.~\ref{fig:verysmallx}, where they are seen to diverge for $x\lessabout 10^{-3}$.  To the extent 
that the $\ln^2(1/x)$ behavior is preferable theoretically, 
we question the reliability at very small $x$ of parton distribution functions based on an assumed 
inverse power behavior.  

The $\ln^2(1/x)$ behavior that we show is consistent with the DIS data may be the signal for the 
onset of a new regime, the physics of saturation or gluon recombination processes $g+g\rightarrow g$ 
at high parton densities.  Thus, new theoretical effort is warranted to devise a QCD evolution 
framework compatible with $\ln^2(1/x)$ behavior of parton densities at 
very small $x$, and experimental programs should be pursued to measure the $x$ and $Q^2$ 
variations of structure functions at much smaller values of $x$ than are currently explored.    

Our next goals include a reanalysis of all available data for $F_2^p(x,Q^2)$ and 
$\partial {F_2^p}_x(x,Q^2)/\partial \ln (Q^2)$ for $x\le x_P$, in $ep$, $\mu p$, and $\nu p$ 
deep inelastic scattering, in order to obtain a joint fit to both the $x$ and $Q^2$ 
dependences, constrained by the Froissart bound and the scaling point.  This work 
should allow us to make more accurate predictions of the proton structure function at 
very small $x$ and very large $Q^2$, regions beyond the reach of existing accelerators.  
We also will investigate the domain of compatibility in Bjorken-$x$  of a $\ln^2 1/x$ 
behavior of $F_2^p(x,Q^2)$ at very small $x$ with quark and gluon distribution functions
obtained in the standard fashion, e.g., in Ref.~\cite{Tung:2006tb},
with DGLAP evolution and an assumed inverse power behavior of PDFs.

\begin{acknowledgments}
E.L.B. is supported by the U.~S.\ Department of Energy, Division of High 
Energy Physics, under Contract No.\ DE-AC02-06CH11357.
C-I.T. is supported in part by the U.~S.~Department of Energy under 
Contract DE-FG02-91ER40688, TASK A.  E.L.B thanks Pavel Nadolksy for his  
assistance in obtaining values of $F_2$ and uncertainties from the CTEQ6.5 
parameterization.  M.M.B. thanks Professor  L. Durand III for very valuable discussions and the Aspen Center for Physics for its 
hospitality during the writing of this paper. 


\end{acknowledgments}


\end{document}